\documentclass[useAMS,usenatbib]{mn2e}
\usepackage[dvips]{color,graphicx}
\usepackage{multirow}
\newcommand{\bi}{\bfseries\itshape} 

\def\aj{AJ}             	% Astronomical Journal
\def\araa{ARA\&A}       	% Annual Review of Astron and Astrophys
\def\apj{ApJ}           	% Astrophysical Journal
\def\apjl{ApJ}          	% Astrophysical Journal, Letters
\def\apjs{ApJS}         	% Astrophysical Journal, Supplement
\def\aap{A\&A}          	% Astronomy and Astrophysics
\def\mnras{MNRAS}       	% Monthly Notices of the RAS
\def\pasj{PASJ}         	% Publications of the ASJ
\def\nat{Nature}        	% Naturerun_fitpsf3d.pro
\def\na{New Astron.}        	% Naturerun_fitpsf3d.pro
\def\jcap{JCAP}   % Astrophysics Letters%%

\def\afe{\rm[\alpha/Fe]}

\def\feh{\rm[Fe/H]}

\def\gsim{\hspace{0.3em}\raisebox{0.4ex}{$>$}\hspace{-0.75em}\raisebox{-.7ex}{$\sim$}\hspace{0.3em}}
\def\lsim{\hspace{0.3em}\raisebox{0.4ex}{$<$}\hspace{-0.75em}\raisebox{-.7ex}{$\sim$}\hspace{0.3em}}
\title[Thick discs in clumpy galaxies]{Properties of thick discs formed in clumpy galaxies}
\author[S. Inoue \& T. R. Saitoh]{Shigeki
Inoue$^{1,2}$\thanks{E-mail: shigeki.inoue@mail.huji.ac.il} \& Takayuki R. Saitoh $^{3}$
\\
$^{1}$Korea Astronomy and Space Science Institute 776, Daedeokdae-ro, Yuseong-gu, Daejeon, 305-348, Republic of Korea\\
$^{2}$Racah Institute of Physics, The Hebrew University, Jerusalem, 91904, Israel\\
$^{3}$Earth-Life Science Institute, Tokyo Institute of
Technology, 2--12--1 Ookayama, Meguro, Tokyo 152--8551, Japan}

\begin{document}

\date{2013 November 11}

\pagerange{\pageref{firstpage}--\pageref{lastpage}} \pubyear{2013}

\maketitle

\label{firstpage}

\begin{abstract}
We examine a possible formation scenario of galactic thick discs with numerical simulations. Thick discs have previously been argued to form in clumpy disc phase in the high-redshift Universe, which host giant clumps of $\hspace{0.3em}\raisebox{0.4ex}{$<$}\hspace{-0.75em}\raisebox{-.7ex}{$\sim$}\hspace{0.3em}10^{9}~{\rm M_\odot}$ in their highly gas-rich discs. We performed $N$-body/smoothed particle hydrodynamics simulations using isolated galaxy models for the purpose of verifying whether dynamical and chemical properties of the thick discs formed in such clumpy galaxies are compatible with observations. The results of our simulations seem nearly consistent with observations in dynamical properties such as radial and vertical density profiles, significant rotation velocity lag with height and distributions of orbital eccentricities. In addition, the thick discs in our simulations indicate nearly exponential dependence of $\sigma_\theta$ and $\sigma_z$ with radius, nearly isothermal kinematics in vertical direction and negligible metallicity gradients in radial and vertical directions. However, our simulations cannot reproduce altitudinal dependence of eccentricities, metallicity relations with eccentricities or rotation velocities, which shows striking discrepancy from recent observations of the Galactic thick disc. From this result, we infer that the clumpy disc scenario for thick-disc formation would not be suitable at least for the Milky Way. Our study, however, cannot reject this scenario for external galaxies if not all galaxies form their thick discs by the same process. In addition, we found that a significant fraction of thick-disc stars forms in giant clumps.

\end{abstract}


\section{Introduction}
\label{Intro}
Duplex structures of thin and thick discs seem to be ubiquitous among disc galaxies \citep[e.g.][]{b:79,db:02,yd:06}, including the Milky Way (MW) \citep[e.g.][]{y:82,gr:83}. Because thick-disc stars are generally old and $\alpha$-enhanced population which implies their early and rapid formation \citep[e.g.][]{abw:06,hdl:13}, structures of the thick discs are expected to reflect galactic formation history. The origin of the thick discs is, however, still an open question, and a number of possible formation scenarios have been proposed so far. Ones frequently discussed in recent studies are as follows: stellar accretion from disrupted satellites \citep[e.g.][]{ans:03}, thickening of a thin disc by minor mergers \citep[e.g.][]{qhf:93,kbz:08}, multiple gas-rich minor mergers \citep{bkg:04,bgm:05,bsg:12}, and radial migration by spiral arms and barred structures in a disc (e.g. \citealt{sb:09b,sb:09a}). \citet{sha:09} have discussed that distribution of orbital eccentricities of thick-disc stars reflects their formation processes and can be used as a probe of these scenarios. Afterwards, some observations were performed for the thick disc of the MW and elucidated that the thick-disc stars have a mono-modal distribution with the peak at the eccentricity of $\simeq0.2$ \citep{dkr:10,lba:11,dgk:11,r:11,krl:11,kgw:13}. They argued that the multiple gas-rich minor mergers would be a favourable scenario for the Galactic thick-disc formation, and the radial migration scenario could also reproduce the eccentricity distribution in acceptable matching. On the other hand, the stellar accretion scenario results in relatively higher eccentricities than the observations, and the thin-disc thickening by dwarf galaxies seems to bring about a bi-modal distribution. In addition, observations of \citet{lba:11} found that correlations between metallicities and some kinematic properties are obviously different between the Galactic thin and thick discs.

In the high-redshift Universe, some galaxies have been observed to be clumpy\footnote{These discy galaxies are often called clump clusters, tadpole and chain galaxies according to their inclinations.} , and they have giant clumps the masses of which are $\hspace{0.3em}\raisebox{0.4ex}{$<$}\hspace{-0.75em}\raisebox{-.7ex}{$\sim$}\hspace{0.3em}10^{9}~{\rm M_\odot}$ in their highly gas-rich discs \citep[e.g.][]{bar:96,eeh:04,fgl:06,g:06,ggf:11,bce:13,tkt:13II}. Numerical simulations demonstrated the evolution from clump clusters to disk galaxies \citep[e.g.][]{n:98,n:99,isg:04,bee:07,atm:09,dsc:09,cdb:10,mdc:13}. In the evolutionary process, giant clumps fall into the galactic centre by dynamical friction and can eventually form a galactic bulge \citep{n:98,n:99,ebe:08,is:12,dk:13,bpr:13,pvt:13}. In addition, the clumps are so massive that they can change dark matter (DM) density profiles (\citealt{ebe:08b,is:11}; see also \citealt{dph:14}) and induce significant net rotation of halo stars and globular clusters around galactic discs \citep{i:13}.

\citet{n:96} has pointed out the possibility of thick-disc formation in such a clumpy phase of galaxy evolution: Orbital motions of giant clumps kinematically can heat up a galactic disc, and disrupted clumps supply stars to the thick disc to some extent. Afterwards, \citet{bem:09} demonstrated it by numerical simulations. Their numerical simulations showed that the clumpy disc scenario can reproduce observed thick discs with exponential density profiles \citep[see also][]{bee:07,es:13} and scale heights independent from radius. Thick discs of external galaxies were indeed observed to have a constant thickness \citep{vs:81,jt:82,sg:90,matthews:00,db:02}. They suggested that the reproducibility of the constant thickness is a characteristic point of this formation scenario; the thin-disc thickening by minor mergers lead to flaring of thick discs in outer disc regions \citep[e.g.][]{kbz:08}. Observations also seemed to show that the disc-like structures in the clumpy galaxies are consistent with thick discs in nearby galaxies in their thickness and age \citep{ee:05,ee:06}, and spatial distributions of clumps are nearly exponential although surface brightness profiles of the discs are clearly not \citep{eev:05}. Thus, it could be said that the thick-disc formation in clumpy galaxies is also a possible scenario and should be examined in detail. However, the properties of the thick discs formed in this scenario have not been compared with observations in the way of \citet{sha:09}.

In this study, we performed several runs of high-resolution $N$-body/smoothed particle hydrodynamics (SPH) simulation using isolated models and study dynamical and chemical properties of thick discs formed in the clumpy disc scenario. Our aim in this paper is to scrutinize the credibility of this thick-disc formation scenario and discuss whether we can verify or discard the scenario. We describe our simulation settings in \S\ref{sim} and analyses of the thick-disc properties in \S\ref{Results}. We present discussion and conclusions in \S\ref{discuss} and \S\ref{conc}.

\section{Simulation}
\label{sim}
The detailed description of our computing schemes has been presented in \citet{is:11,is:12}, therefore we describe it briefly here. We employed an $N$-body/SPH code, {\tt ASURA} \citep{sdk:08,sdk:09} using the standard scheme of \citet{m:92} and \citet{s:10}.\footnote{Although the latest version of {\tt ASURA} uses a density-independent SPH scheme \citep{sm:13} and a symmetric form of the Plummer softening \citep{sm:12}, the simulations in this study were performed with the standard SPH and the ordinary softening.} The code adopts the time-step limiter of \citet{sm:09} so that we can handle strong shocks with an individual time-step method \citep{m:86,makino:91}. It also employs the FAST method \citep{sm:10} which accelerates the simulations by using different time-steps for the gravitational and hydrodynamical parts. Gravity is solved by a parallel tree with GRAPE (GRAvity PipE) method \citep{m:04,tyn:13}. 

A metallicity-dependent cooling function of gas assumes optically thin radiative cooling and covers a temperature range of $10-10^8$ K \citep{wps:09}. Feedback from a uniform far-ultraviolet radiation was taken into account. A gas particle spawns a stellar particle the mass of which is one-third of the initial gas particle under the criteria: 1) $\rho_{\rm gas}>{\rm100~atm~cm^{-3}}$, 2) temperature $T<100$ K, 3) $\nabla\cdot${\bi v}$<0$ and 4) there are no SNe around the particle. The local star formation rate follows the Schmidt law \citep{s:59}. We assumed the Salpeter initial mass function \citep{s:55} ranging from $0.1~{\rm M_\odot}$ to $100~{\rm M_\odot}$ on a stellar particle and that stars heavier than 8 M$_\odot$ cause type-II SNe, return their masses to ambient gas and inject thermal energy into it at the rate of $10^{51}~{\rm erg}$ per SN. The initially primordial gas is contaminated with heavy elements released from the SNe. Our simulation code tracks only the total metal abundance, $Z$, and does not follow the evolution of each element. We did not take type-Ia (or other) SNe or stellar wind from intermediate mass stars into account in our simulation.

Our initial condition followed the spherical models of \citet{kmw:06,kmw:07} and \citet{tpd:13} that have been used to study the formation of disc galaxies in isolated environments. We assumed equilibrium systems with the Navarro-Frenk-White profile \citep[][hereafter NFW]{nfw:97} with a virial mass of $M_{\rm vir} = 5.0\times10^{11}~{\rm M_\odot}$, a virial radius of $r_{\rm vir} = 1.67\times10^2$ kpc and a concentration parameter of $c=6.0$. Baryon mass fractions of the systems are set to $0.06$. The baryonic components are initially primordial (zero-metal) gas with virial temperature and follows the same density profile as the DM. We executed four runs of simulations with different initial conditions for the gas spheres: two spin parameters of $\lambda=0.04$ and $0.1$, two types of angular momentum distributions (see below). The choice of $\lambda=0.04$ is motivated by the averaged spin parameter of DM haloes in pure $N$-body cosmological simulations \citep[e.g.][]{bdk:01,imp:11}. However, observational measurements of spin parameters of high-redshift galaxies are still challenging and debated; for instance, \citet{dbf:11} advocated consistency with the results of the pure DM simulations, whereas \citet{bcd:07} proposed $\lambda\simeq0.1$ on average for star-forming galaxies in the high-redshift Universe \citep[see also][]{kds:11}. Hence, we adopt this value as another choice of the initial spin parameters. As for the angular momentum distributions, the pure $N$-body simulations indicated specific angular momenta of haloes follow the profile of $j\propto r$ in all radial ranges \citep{bdk:01}. On the other hand, \citet{kds:11} performed cosmological simulations including baryonic physics and showed that specific angular momentum distributions of gas fallow those of DM inside $0.1r_{\rm vir}$ and are constant outside $0.1r_{\rm vir}$; moreover, they found that this distribution is almost independent of redshift. Therefore, we take these two cases as initial conditions of angular momentum distributions. Table \ref{ini_con} summarises our initial settings of the gas spheres.\footnote{The simulation of B01-$\lambda$010 is the same as in \citet{is:11,is:12}.} Although some of them might be unrealistic, we show that properties of the resultant galaxies are qualitatively consistent in all runs (see \S3). The DM and the gas are represented by $10^7$ and $5.0\times10^6$ particles, i.e. the masses of the DM and gas particles are $4.7\times10^4M_\odot$ and $6.0\times10^3M_\odot$. Softening lengths are $10~{\rm pc}$ for all particles.\footnote{Only in B01-$\lambda$010, we used softening lengths of $8~{\rm pc}$ and $2~{\rm pc}$ for the DM and gas particles; however, we confirmed that the difference of the softening lengths does not change our results.}

\begin{table}
  \caption{Summery of initial conditions of our simulations. From left to right, the columns are simulation names, spin parameters and radial distributions of specific angular momentum. The other parameters are the same in all runs.}
  \label{ini_con}
  $$ 
  \begin{tabular}{cccc}
    \hline
    run  & $\lambda$ & $j(r)$ \\
    \hline
    B01-$\lambda$004 & 0.04 & $j\propto r$ \\
    B01-$\lambda$010 & 0.1  & $j\propto r$ \\
    \multirow{2}{*}{K11-$\lambda$004} & \multirow{2}{*}{0.04} & $j\propto r$ in $r<0.1r_{\rm vir}$,\\
                                      &                       & $j=j(0.1r_{\rm vir})$ in $r>0.1r_{\rm vir}$\\
    \multirow{2}{*}{K11-$\lambda$010} & \multirow{2}{*}{0.1} & $j\propto r$ in $r<0.1r_{\rm vir}$,\\
                                      &                       & $j=j(0.1r_{\rm vir})$ in $r>0.1r_{\rm vir}$\\
    \hline
  \end{tabular}
  $$ 
\end{table}

\section{Results}
\label{Results}

\begin{figure*}
  \begin{minipage}{\hsize}
    \includegraphics[width=\hsize]{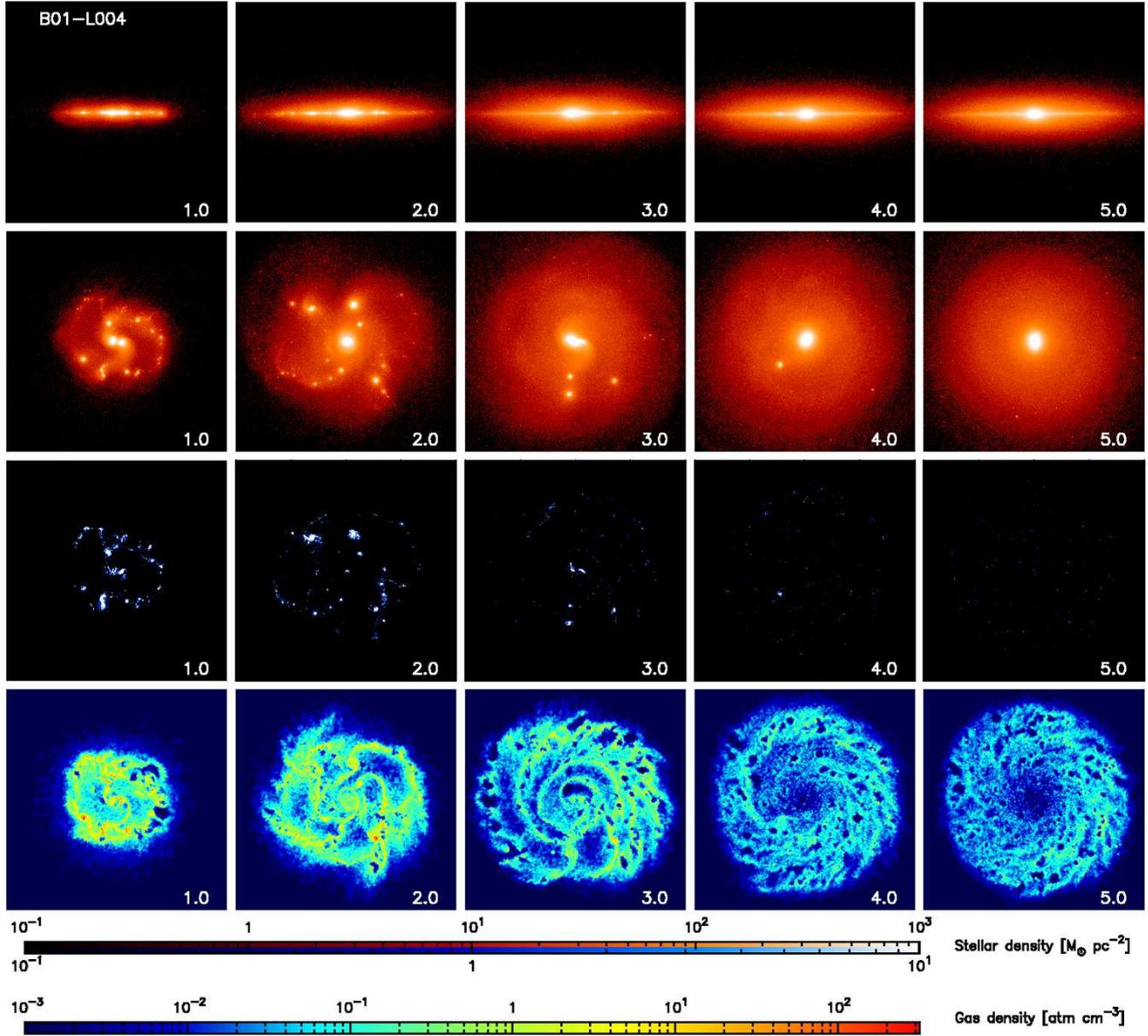}
   \caption{Stellar and gas density maps in the central $40\times40$ kpc region in the run of B01-$\lambda$004. The first and second rows indicate the surface density of all stars from the edge-on and face-on views, respectively. The third row plots the density of stars younger than 40 Myr as star forming region. The fourth row shows the volume density of gas on the disc plane. Time in units of Gyr is indicated on right bottom corner of each panel.}
   \label{snap}
  \end{minipage}
\end{figure*}
\begin{figure*}
  \begin{minipage}{\hsize}
    \includegraphics[width=\hsize]{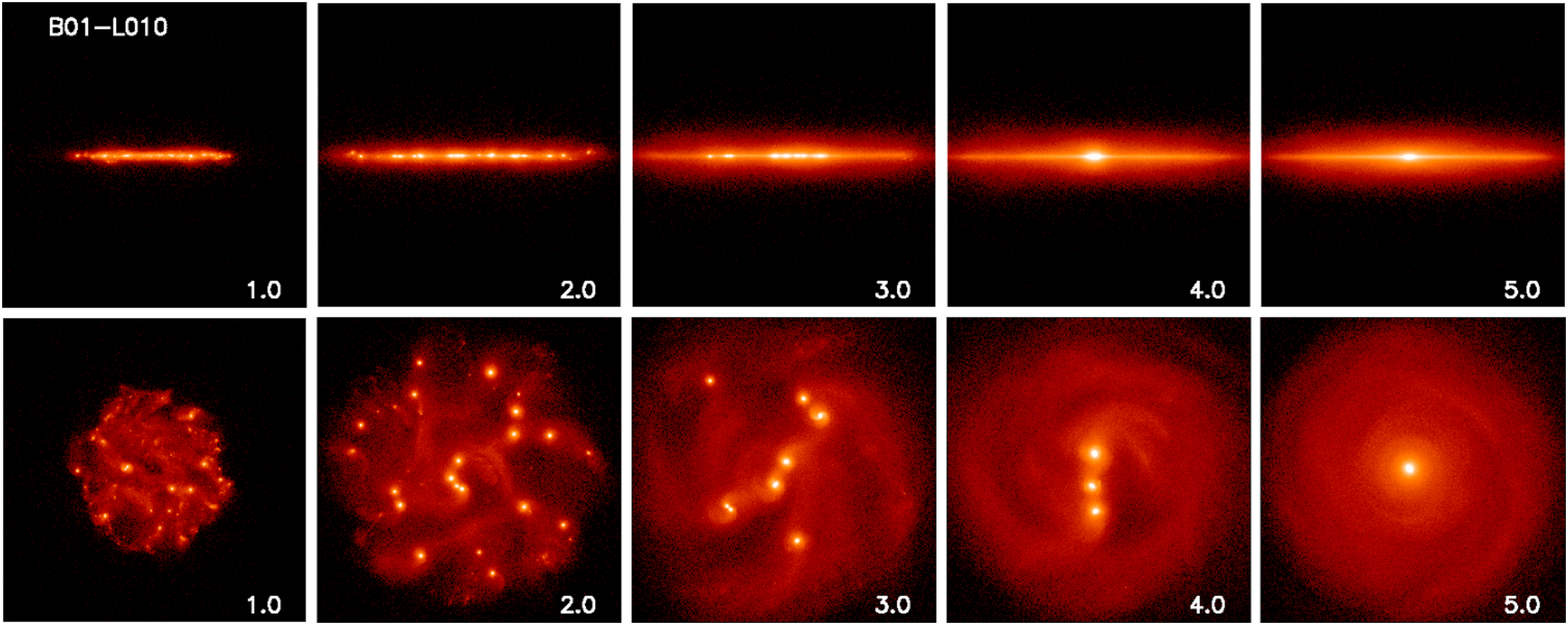}
    \includegraphics[width=\hsize]{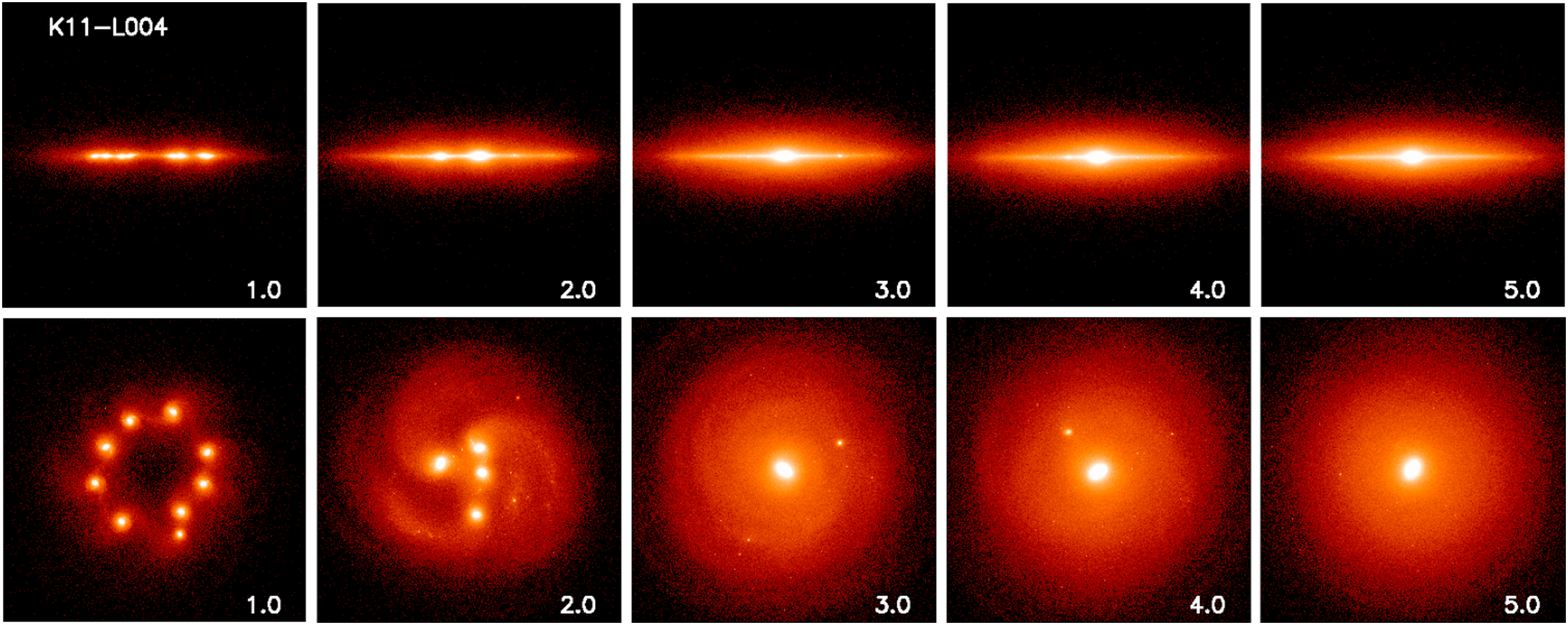}
    \includegraphics[width=\hsize]{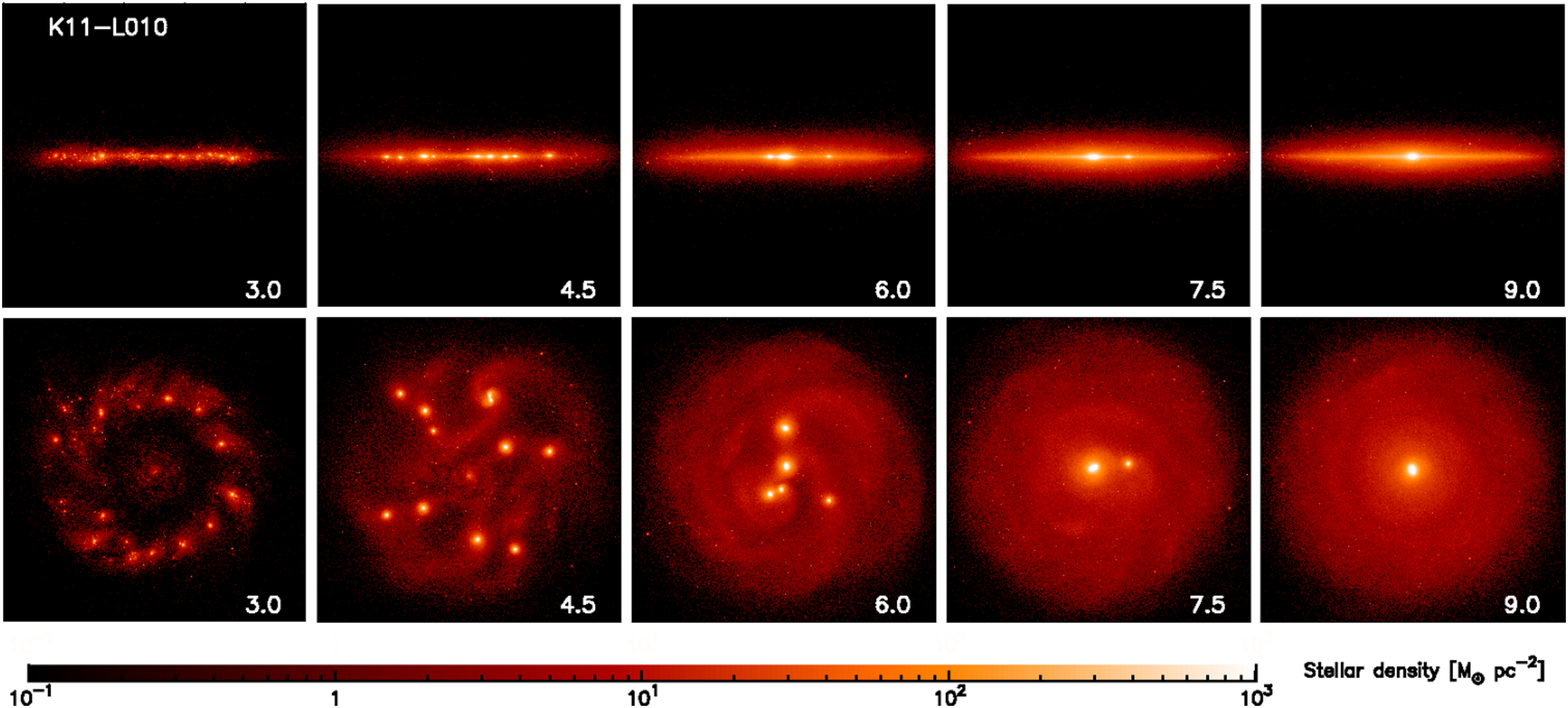}
   \caption{Stellar density maps seen edge-on and face-on in the runs of B01-$\lambda$010, K11-$\lambda$004 and K11-$\lambda$010 from top to bottom. The depicted regions are the central $40\times40$ kpc region for K11-$\lambda$004 (the third and forth rows) and $60\times60$ kpc regions for B01-$\lambda$010 (the first and second rows) and K11-$\lambda$010 (the fifth and sixth rows). Time in units of Gyr is indicated on right bottom corner of each panel.}
   \label{snap_others}
  \end{minipage}
\end{figure*}

Fig. \ref{snap} displays snapshots of stellar and gas distributions and star-forming regions in the run of B01-$\lambda$004, where the clumpy phase of the disc can be seen in $t\lsim2~{\rm Gyr}$. Fig. \ref{snap_others} shows stellar density maps of the other runs. In K11-$\lambda$004 and K11-$\lambda$010, the formation of clumps may be involved with `ring' instability caused by the initial setting of the angular momentum distributions \citep{gm:83,cw:03}. We run our simulations until $t=6~{\rm Gyr}$ except the run of K11-$\lambda$010. Because this run took $\sim8~{\rm Gyr}$ to settle into a stable state, we continued the run until $t=10~{\rm Gyr}$; in this sense, K11-$\lambda$010 may not be a realistic initial condition since the clumpy phase is too long in comparison to other numerical simulations \citep{bee:07,bpr:13,be:09,dsc:09,cdb:10,cdm:11,dk:13,pvt:13}. In the following subsections, we analyse the final state of the galaxies in our simulations.

\subsection{Density profiles}
\label{DensityProfiles}
\subsubsection{Radial profiles}
\label{RadialProfiles}

\begin{figure}
  \includegraphics[width=\hsize]{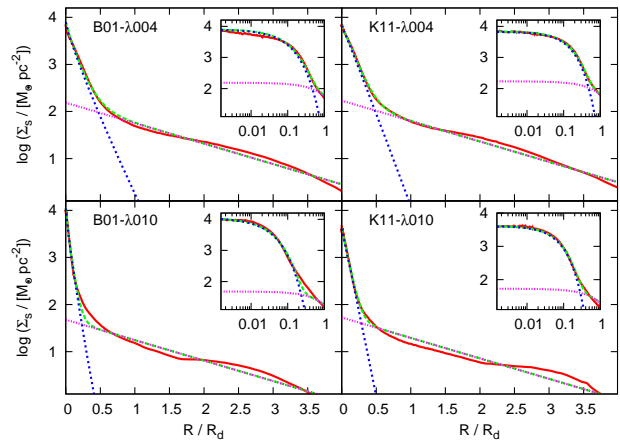}
  \caption{Stellar surface density profiles seen face-on in the final state of our simulations (red lines). Minimum $\chi^2$ fittings are also plotted: S\'ersic profiles for bulges (blue short-dashed lines), the exponential profiles for discs (pink dotted lines) and disc$+$bulge (green long-dashed lines). The insert diagrams illustrate the same profiles with the logarithmic abscissa. The centres of the systems are defined to be the positions at which the stellar densities become the highest.}
  \label{densities}
\end{figure}

\begin{table*}
  \begin{center}
    \begin{minipage}{150mm}
      \caption{The best-fit parameters for the radial profiles. Surface density profiles of discs and bulges are fitted with equation (\ref{sersic}), and $M_{\rm d}$ and $M_{\rm b}$ are disc and bulge masses computed by integrating the fitting profiles. $f_{\rm gas}$ is a mass fraction of gas particles to the total baryon in the region of $R<5R_{\rm d}$ and $|z|<5h_{\rm T}$ in the final state (see equation (\ref{doubleExp}) and Table \ref{bestfit_v} for $h_{\rm T}$).}
      \label{bestfit}
      $$ 
      \begin{tabular}{ccccccccccc}
        \hline
        \noalign{\smallskip}
        \multirow{2}{*}{run} & $\Sigma_{\rm d}$ & $R_{\rm d}$ & $\Sigma_{\rm b}$ & $R_{\rm b}$ & \multirow{2}{*}{$n_{\rm b}$} & $M_{\rm d}$ & $M_{\rm b}$ & \multirow{2}{*}{$f_{\rm gas}$} \\
        & ${\rm [M_\odot~pc^{-2}]}$ & ${\rm [kpc]}$ & ${\rm [M_\odot~pc^{-2}]}$ & ${\rm [kpc]}$ & & ${\rm [M_\odot]}$ & ${\rm [M_\odot]}$  &\\
        \hline
        B01-$\lambda$004 & $1.5\times10^2$ & $4.4$ & $7.8\times10^3$ & $0.38$ & $1.16$ & $1.9\times10^{10}$ & $9.6\times10^9$ & $0.07$ \\
        B01-$\lambda$010 & $4.8\times10$   & $8.1$ & $1.1\times10^4$ & $0.24$ & $1.19$ & $2.0\times10^{10}$ & $5.8\times10^9$ & $0.16$ \\
        K11-$\lambda$004 & $1.7\times10^2$ & $4.1$ & $6.7\times10^3$ & $0.50$ & $0.97$ & $1.8\times10^{10}$ & $9.9\times10^9$ & $0.04$ \\
        K11-$\lambda$010 & $5.3\times10$   & $6.8$ & $4.1\times10^3$ & $0.44$ & $0.98$ & $1.6\times10^{10}$ & $4.7\times10^9$ & $0.29$ \\
        \hline
      \end{tabular}
      $$ 
    \end{minipage}
  \end{center}
\end{table*}
To begin with, we decompose a stellar surface density profile of the simulated galaxy into an exponential disc and a S\'ersic bulge,
\begin{equation}
\label{sersic}
\Sigma_{\rm s}(R) = \Sigma_{\rm d}\exp\left(-\frac{R}{R_{\rm d}}\right) + \Sigma_{\rm b}\exp\left[-\left(\frac{R}{R_{\rm b}}\right)^\frac{1}{n_{\rm b}}\right],
\end{equation}
where $\Sigma_{\rm d,b}$ and $R_{\rm d,b}$ are central surface densities and scale radii of the disc and bulge, and $n_{\rm b}$ is a S\'ersic index of the bulge. The profiles are shown in Fig. \ref{densities}, and the best-fit parameters and the integrated disc and bulge masses are tabulated in Table \ref{bestfit}. The discs of B01-$\lambda$010 and K11-$\lambda$010 have large scale lengths of $\gsim7~{\rm kpc}$; in this sense, B01-$\lambda$010 and K11-$\lambda$010, i.e. the large spin parameter of $\lambda=0.1$, may be unsuitable for reproducing MW-like galaxies. Although the disc scale lengths of B01-$\lambda$004 and K11-$\lambda$004 may still be a little larger than that of the MW in recent observations \citep[e.g.][]{jib:08,bab:12II,br:13} or within a range of uncertainty \citep[e.g.][]{crm:12}, these are compatible with typical values of disc galaxies \citep[e.g.][]{kent:85,bm:98,fab:10,f:10}.

In all runs, both the discs and the bulges can be fitted well with exponential profiles: the S\'ersic indices of the bulges are $\simeq1$. Such small S\'ersic indices mean that the bulges should be classified into pseudo-bulges \citep[e.g.][]{kk:04,df:07,fd:08}. However, other studies using numerical simulations with isolated models have supported the formation of classical bulges with high S\'ersic indices as remnants of the clump coalescence \citep{ebe:08,pvt:13}. In addition, cosmological simulations also seem to indicate $n_{\rm b}=3$--$5$ (D. Ceverino et al. in prep.). The diversity of the S\'ersic indices between these studies and ours may be related to gas fraction of the clumps. If the progenitor clumps and the remnant bulges are gas-rich, the dissipative gas can contract to the centre and may form highly concentrated density profiles, which leads to a high S\'ersic index. On the other hand, an opposite expectation may also be possible. High S\'ersic indices can be considered as results of violent relaxation of dissipationless mergers like elliptical galaxies. In this case, coalescence of gas-poor clumps may lead to a high S\'ersic index, rather than gas-rich clumps. Thus, the dynamical properties of `clump-origin bulges' are to be investigated further in future studies.\footnote{Detailed analysis of the bulge in B01-$\lambda$010 has been reported in \citet{is:12}.}

\subsubsection{Vertical profiles}
\label{VerticalProfiles}
\begin{figure}
  \includegraphics[width=\hsize]{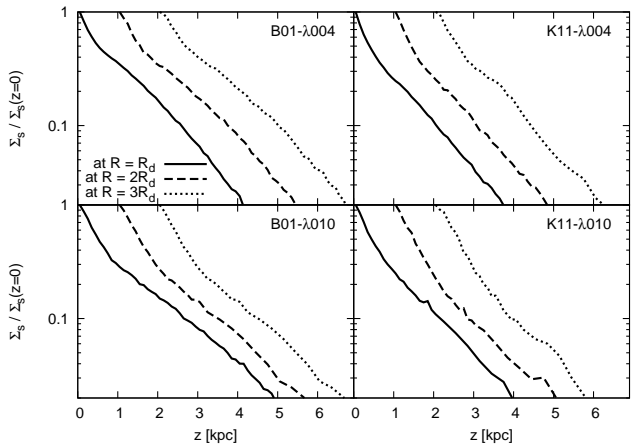}
  \caption{Vertical density profiles of stars in projected edge-on view at the final state of the simulations. The profiles are normalized by the surface density at $z=0$ and shifted right at intervals of $1~{\rm kpc}$ for visibility.}
  \label{VertDen}
\end{figure}

\begin{table*}
  \begin{center}
    \begin{minipage}{160mm}
      \caption{The best-fit parameters for vertical density profiles with equation (\ref{doubleExp}). Subscriptions of `T' and `t' mean values for thick and thin discs.}
      \label{bestfit_v}
      $$ 
      \begin{tabular}{cccccccccc}
        \hline
        \noalign{\smallskip}
        \multirow{2}{*}{run} & $\rho_{0, \rm T}$ & $R_{\rm T}$ & $h_{\rm T}$ & $M_{\rm d, T}$ & $\rho_{0, \rm t}$ & $R_{\rm t}$ & $h_{\rm t}$ & $M_{\rm d, t}$ & \multirow{2}{*}{$\frac{M_{\rm d, T}}{M_{\rm d, t}}$} \\
        & ${\rm [M_\odot~pc^{-3}]}$ & ${\rm [kpc]}$ & ${\rm [kpc]}$ & ${\rm [M_\odot]}$ & ${\rm [M_\odot~pc^{-3}]}$ & ${\rm [kpc]}$ & ${\rm [kpc]}$ & ${\rm [M_\odot]}$ &\\
        \hline
        B01-$\lambda$004 & $3.5\times10^{-2}$ & $5.5$ & $1.4$ & $2.0\times10^{10}$ & $2.7\times10^{-2}$ & $5.5$ & $0.19$ & $2.0\times10^9$ & $10.0$ \\
        B01-$\lambda$010 & $6.5\times10^{-3}$ & $9.8$ & $1.9$ & $1.5\times10^{10}$ & $1.4\times10^{-2}$ & $9.8$ & $0.45$ & $7.4\times10^9$ & $2.1$ \\
        K11-$\lambda$004 & $7.3\times10^{-2}$ & $4.2$ & $1.1$ & $1.8\times10^{10}$ & $6.5\times10^{-2}$ & $4.2$ & $0.17$ & $2.5\times10^9$ & $7.2$ \\
        K11-$\lambda$010 & $1.0\times10^{-2}$ & $8.2$ & $1.3$ & $1.1\times10^{10}$ & $1.3\times10^{-2}$ & $8.2$ & $0.31$ & $3.5\times10^9$ & $3.2$ \\
        \hline
      \end{tabular}
      $$
    \end{minipage}
  \end{center}
\end{table*}

Fig. \ref{VertDen} shows vertical density profiles of the projected stellar discs at different galactocentric radii. The profiles can be fitted with superposition of two exponential profiles with breaks at $z\simeq1~{\rm kpc}$: thin and thick discs. In agreement with the results of \citet{bem:09}, vertical scale heights of the thick discs are almost independent from radius in all of our simulations; we found that the variation of the scale heights were smaller that $20$ per cent in the range of $1\leq R/R_{\rm d}\leq3$. In addition, scale radii and central densities of the thin and thick discs can be measured by the fitting of the vertical profiles with double-exponential models,
\begin{equation}
  \rho_{\rm d}(R,z) = \sum_i\rho_{0,i}\exp\left(-\frac{R}{R_{i}}-\frac{|z|}{h_i}\right),
  \label{doubleExp}
\end{equation}
where the subscript $i$ takes thin and thick discs, and $R_i$ and $h_i$ are scale radii and scale heights of the discs. The fitting is performed in the range of $1\leq R/R_{\rm d}\leq3$ to avoid the bulge contribution. We found that the scale radii are hardly different between the thin and thick discs. The best-fit parameters and masses of the thin and thick discs are tabulated in Table \ref{bestfit_v}.

In all runs, however, the thick discs are more massive than the thin discs; we found mass ratios of the thick to the thin discs to be greater than $2$ in all runs. Thick discs have been observed to account for $5$--$40$ per cent of the total stellar masses in real galaxies \citep{yd:06}. Although \citet{f:08} and \citet{cek:11} suggested that typically thick and thin discs have similar masses, the thin discs in our simulations would be still too less massive. This is because our initial conditions assumed single collapse models, do not include late accretion to feed gas to the thin discs after the clumpy disc formation \citep{bem:09}. We warn the reader that this point is a limitation of our simulations. It should be noted that the thick-disc scale heights would be somewhat shortened by the thin-disc formation that is lacked in our simulations \citep{bem:09}. Moreover, the scale radii of the thin discs may become shorter or longer than those of their thick discs after the completion of the thin-disc formation.

\subsection{Kinematics}
\subsubsection{Rotation velocities}
\begin{figure}
  \includegraphics[width=\hsize]{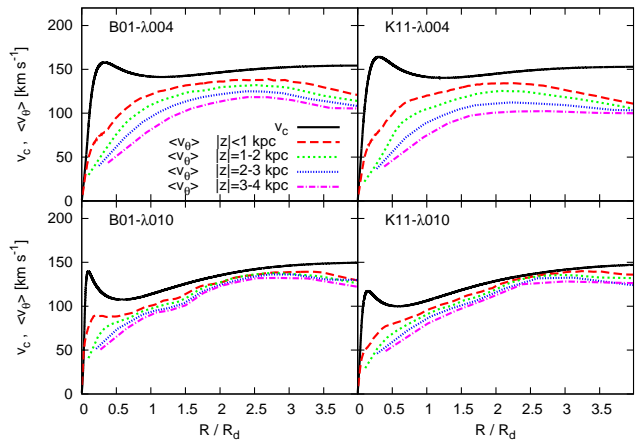}
  \caption{Radial profiles of circular and mean rotation velocities of stars. The circular velocity is computed as $v_c=\sqrt{GM(<R)/R}$, where $G$ is the gravitational constant and $M(<R)$ is the mass of the system spherically enclosed in radius of $R$. The mean rotation velocity $v_\theta$ is measured at each height above the plane.}
  \label{CircV}
\end{figure}
Fig. \ref{CircV} shows radial profiles of circular and mean rotation velocities of stars. The galaxies have nearly flat rotation curves in the runs of B01-$\lambda$004 and K11-$\lambda$004 in their disc regions, whereas the circular velocities gently increase in B01-$\lambda$010 and K11-$\lambda$010 (see also Fig. \ref{LeftRight}). It should be noted, however, that the massive bulges cause the mass concentrations of the galaxies to be high as seen in Fig. \ref{densities}, and contribution of the bulges to the circular velocity curves would be substantially present in all simulated runs.

The mean rotation velocities in the figure show significant asymmetric drift in the cases of B01-$\lambda$004 and K11-$\lambda$004: velocity lags of $\mathrm{d}\overline{v_\theta}/\mathrm{d}z\sim-10~{\rm km~s^{-1}~kpc^{-1}}$. In B01-$\lambda$010 and K11-$\lambda$010, however, the velocity lags are weak. In the MW, rotation velocity has been observed to decrease significantly with distance from the plane although determinations do not seem be consistent quantitatively: from $-36$ to $-9~{\rm km~s^{-1}~kpc^{-1}}$ \citep{cb:00,cbc:10,slf:10,lba:11,dgk:11,mcm:12}; in this sense, B01-$\lambda$004 and K11-$\lambda$004 seem to be more similar to the MW. Meanwhile, in external galaxies, \citet{yd:08} have observed that massive galaxies in their sample indicate little differences of rotation velocities between thin and thick discs.

\subsubsection{Velocity dispersions}
\begin{figure}
  \includegraphics[width=\hsize]{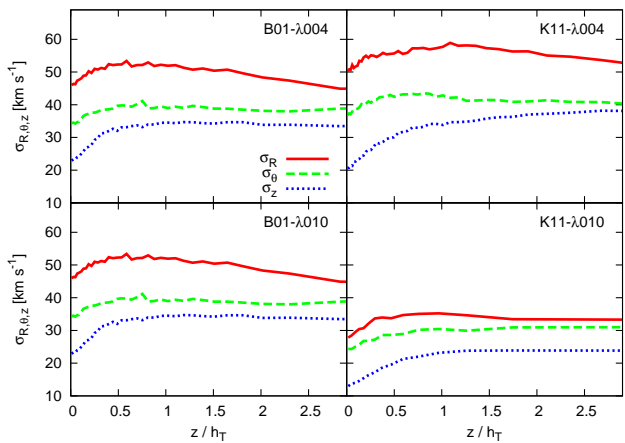}
  \caption{Dispersions of radial, rotation and vertical velocities as functions of distance from the plane at the galactocentric radius of $R=2R_{\rm d}$.}
  \label{VertDisp}
\end{figure}
Fig. \ref{VertDisp} shows vertical profiles of velocity dispersions. As seen from the figure, the velocity dispersions are almost constant, and the discs can be considered to be isothermal along the vertical direction. Hereafter, we follow the analytical discussion of \citet{lf:89}. If the surface mass density is dominated by the stellar disc and the vertical density profile can be assumed to be a single exponential function, from the Jeans equation with the isothermality, 
\begin{equation}
  \sigma_z^2(R) \propto \Sigma_{\rm s} \propto \exp\left(-\frac{R}{R_{\rm d}}\right)
\end{equation}
Furthermore, if we assume uniform velocity anisotropy throughout the discs: $\sigma_\theta/\sigma_z\simeq {\rm const}$, $\sigma_\theta^2(R)\propto \exp\left(-R/R_{\rm d}\right)$. Finally, if $\sigma_R^2$ is exponential with a scale length of $R_R$,
\begin{equation}
  \frac{\sigma_\theta^2}{\sigma_R^2} = \frac{\sigma_\theta^2(0)}{\sigma_R^2(0)}\exp\left[-R\left(\frac{1}{R_{\rm d}}-\frac{1}{R_R}\right)\right] = \frac{1}{2}\left(1+\frac{R}{v_{\rm c}}\frac{\mathrm{d}v_{\rm c}}{\mathrm{d}R}\right),
  \label{flat_rotation}
\end{equation}
where the second equality is derived from the epicyclic approximation \citep[e.g.][]{bt:08}. In the case of a flat rotation curve, since the right-hand side (RHS) is constant, the two scale lengths must be $R_R=R_{\rm d}$; moreover, $\sigma_\theta^2/\sigma_R^2=1/2$. To summarise, if all of these assumptions hold, it can be expected that
\begin{equation}
\Sigma_{\rm s} \propto \sigma_R^2 \propto \sigma_\theta^2 \propto \sigma_z^2 \propto \exp\left(-\frac{R}{R_{\rm d}}\right).
\end{equation}

\begin{figure}
  \includegraphics[width=\hsize]{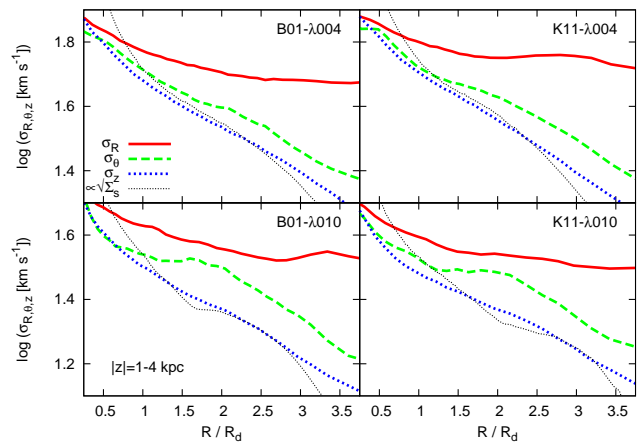}
  \caption{Radial, rotation and vertical velocity dispersions as functions of galactocentric distance. Stars in $1~{\rm kpc}<|z|<4~{\rm kpc}$ are taken into account for the measurements. For the sake of comparison, normalized profiles of the square roots of stellar surface densities are also plotted with the black thin dotted lines.}
  \label{RadDisp}
\end{figure}
\begin{figure}
  \includegraphics[width=\hsize]{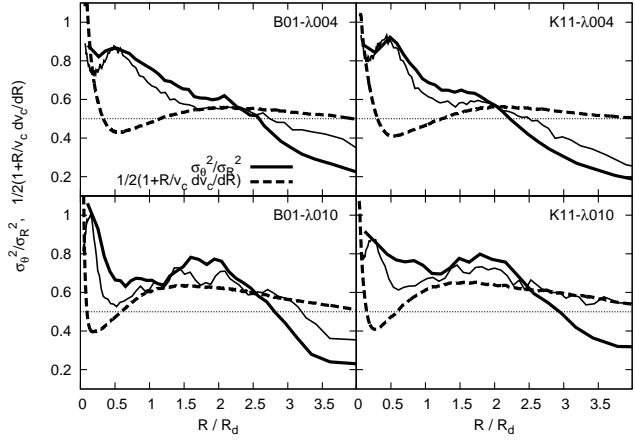}
  \caption{The left- and right-hand sides of equation (\ref{flat_rotation}) as functions of radius. The thin and thick lines indicate $\sigma_\theta^2/\sigma_R^2$ for stars in the regions of $|z|<1~{\rm kpc}$ and $1~{\rm kpc}<|z|<4~{\rm kpc}$, respectively. The horizontal dotted-lines indicate 1/2.}
  \label{LeftRight}
\end{figure}
To compare our simulations with the above analyses, Fig. \ref{RadDisp} shows radial profiles of velocity dispersions of thick-disc stars in our simulations. We confirmed that the profiles hardly depend on height above the plane because of the isothermality shown in Fig. \ref{VertDisp}. As discussed above, the profiles of $\sigma_\theta$, $\sigma_z$ and $\sqrt{\Sigma_{\rm s}}$ seem to indicate exponential behaviour with nearly similar scale lengths although $\sigma_\theta$ have shallower slopes in $1<R/R_{\rm d}<2$ in B01-$\lambda$010 and K11-$\lambda$010. The decrease of $\sigma_R$ with radius, however, are much slower than the other profiles and not exponential. This means that the assumption of uniform velocity anisotropy between $\sigma_R$ and the other dispersions is not valid in the simulated discs. Fig. \ref{LeftRight} shows the left- and RHS of equation (\ref{flat_rotation}). The ratios of $\sigma_\theta^2/\sigma_R^2$ are not constant and decrease with radius in outer regions although the RHS is $\simeq1/2$, indicating that the circular velocity curves can be considered to be almost flat. It is particularly worth noting that $\sigma_\theta^2/\sigma_R^2$ largely decrease and deviate from the RHS in $R\gsim2.5R_{\rm d}$ except the region of $|z|<1~{\rm kpc}$ in K11-$\lambda$010. This implies that the epicyclic approximation would not be valid in the outer disc regions in our simulations.
%the right-hand side of equation (\ref{flat_rotation}) is expected to increase with radius. Notwithstanding, the left-hand side, $\sigma_\theta^2/\sigma_R^2$, clearly decreases in all runs. Accordingly, we can see that the analysis of equation (\ref{flat_rotation}) is not  applicable to the simulation results. 
Another notable point in Fig. \ref{RadDisp} is that the behaviour of the velocity dispersions are almost the same among the simulation runs.

\subsubsection{Distributions of orbital eccentricities}
\label{Eccentricity_disp}
As we mentioned in \S\ref{Intro}, \citet{sha:09} have found that distribution of orbital eccentricities of thick-disc stars reflects their formation process. Therefore, it is worthwhile to inquire into eccentricity distributions in our simulations. Following the manner of \citet{sha:09}, we fitted the circular velocity curves of Fig. \ref{CircV} with combined models of an NFW halo, a Hernquist bulge \citep{h:90} and a Miyamoto-Nagai disc \citep{my:75}. Orbit calculations are executed in the modeled potentials, and peri- and apo-centre distances $r_{\rm pe, ap}$ are computed, in which the final states of our simulations are used as the initial positions and velocities of the orbit calculations. The orbital eccentricity is defined as $\epsilon\equiv\left(r_{\rm ap}-r_{\rm pe}\right)/\left(r_{\rm ap}+r_{\rm pe}\right)$.
\begin{figure}
  \includegraphics[width=\hsize]{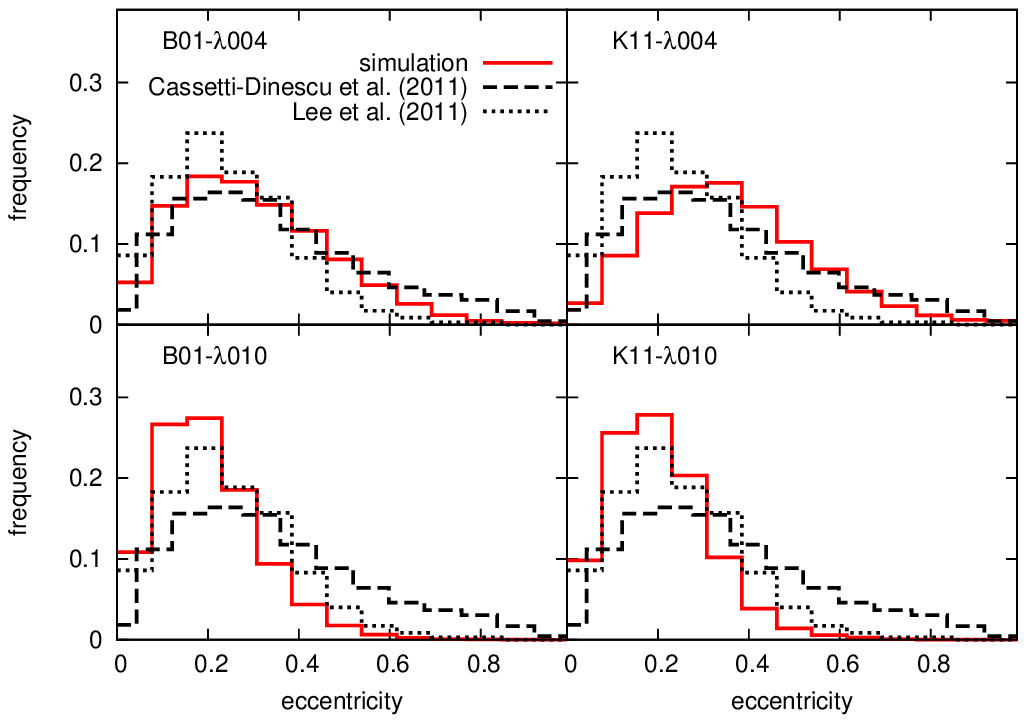}
  \includegraphics[width=\hsize]{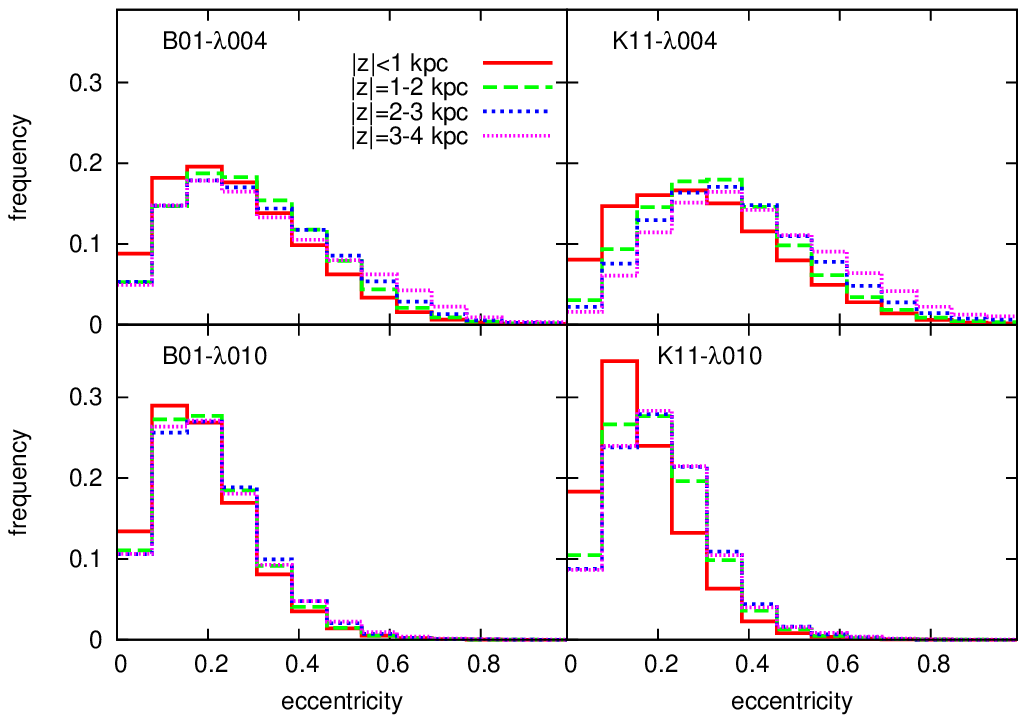}
  \caption{Top: Eccentricity distributions of stars in the region of $2<R/R_{\rm d}<3$ and $1~{\rm kpc}<|z|<4~{\rm kpc}$ in the final states. Observational results by \citet[][in $0.8~{\rm kpc}<|z|<2.4~{\rm kpc}$]{lba:11} and \citet[][in $1~{\rm kpc}<|z|<3~{\rm kpc}$]{dgk:11} are overplotted for comparison. Bottom: Eccentricity distributions at different heights in our simulations.}
  \label{Ecce}
\end{figure}

The top panel of Fig. \ref{Ecce} shows the eccentricity distributions in thick-disc regions of $1~{\rm kpc}<|z|<4~{\rm kpc}$. These distributions are almost consistent with the observations of the Galactic thick disc. The distributions do not indicate significant dependence on the initial conditions although the peak values are high in B01-$\lambda$010 and K11-$\lambda$010 and the position of the peak tends to be a little high eccentricity in K11-$\lambda$004. The bottom panel of Fig. \ref{Ecce} shows dependence of the distributions on distance from the planes. Although the lowest regions of $|z|<1~{\rm kpc}$ may have slightly low eccentricities (see below), the thick-disc stars in our simulations do not indicate altitudinal gradients of orbital eccentricities. It is noteworthy that this would be discrepant from observations: \citet{dkr:10} observed that mean eccentricities decrease with height in the Galactic thick disc, and their observed distributions become broader in higher regions. We discuss the dependence of this result on the galaxy models in Appendix \ref{App} and find that the eccentricity distributions are qualitatively independent from the details of the galaxy models.

From integration of the fitting profiles of vertical density distributions, thin disc stars occupy stellar mass fractions of $0.17$, $0.52$, $0.19$ and $0.36$ in B01-$\lambda$004, B01-$\lambda$010, K11-$\lambda$004 and K11-$\lambda$010 in the lowest regions of $|z|<1~{\rm kpc}$. The slightly lower eccentricities in this regions may be due to the contamination of the thin-disc stars if thin-disc stars have more regular kinematics than thick-disc stars.

\subsection{Metallicity relations}
\label{met_rel}
\begin{figure}
  \includegraphics[width=\hsize]{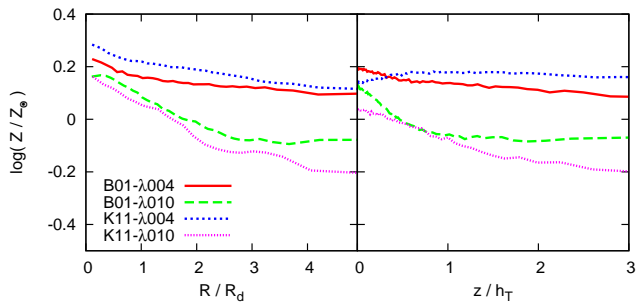}
  \caption{Radial and vertical distributions of mean metallicities of stars. The radial (the right panel) and vertical profiles (the left panel) are measured in $1~{\rm kpc}<|z|<4~{\rm kpc}$ and $2<R/R_{\rm d}<3$, respectively. We set the solar metallicity to $Z_\odot=0.02$.}
  \label{ChemProf}
\end{figure}
Metallicity distribution can also be a key to investigate the formation history of the thick-disc stars. Fig. \ref{ChemProf} shows radial and vertical distributions of averaged metallicity in our simulations.\footnote{All averaged values are mass-weighted throughout this paper, which are hardly different from the simple mean values of stellar particles.} In all runs, stars in the thick-disc regions have metallicities similar to the solar value and are more metal-rich than the observed thick-disc stars \citep[e.g.][and reference therein]{kf:11}. This can be attributed to metal-enrichment caused inside the giant clumps since a significant fraction of thick-disc stars are born in massive clumps in our simulations (see \S\ref{ClumpOriginMass}). Thus, the existence of giant clumps during the disc-formation epoch can lead to significantly high metallicity of thick-disc stars; however, it should be noted that the metal-enrichment would depend on the intensity of feedback in the simulations since gas outflow from the clumps could refresh the gas inside them (see \S\ref{discuss_sim}). Besides, the metallicity gradients in our simulations are weak in both radial and vertical directions. This would be because perturbations of massive clumps vehemently stir the disc stars and wash out their metallicity gradients which could have appeared in quiet disc formation. Although the runs of B01-$\lambda$010 and K11-$\lambda$010 show metallicity gradients in $R\lsim2R_{\rm d}$, the radial variations in metallicities are as small as $\sim0.3~{\rm dex}$. Thus, we expect that the clumpy disc formation would lead to the chemical uniformity of a thick disc if our single collapse models well represent the formation of clumpy galaxies in the high-redshift Universe. The MW thick disc has been observed to have only weak radial and vertical gradients: e.g. $\partial Z/\partial R\simeq-0.003\pm0.005~{\rm dex~kpc^{-1}}$ \citep{hhb:13}, and $\partial Z/\partial z\simeq-0.1\pm0.05~{\rm dex~kpc^{-1}}$ \citep{krl:11,kgw:13} and $-0.26\pm0.02~{\rm dex~kpc^{-1}}$ \citep{hhb:13}. Radial and vertical gradients of $\feh$ are observed to be between $0.0028$ and $0.041~{\rm dex~kpc^{-1}}$ \citep{nma:04,r:11,cab:12,crmm:12,ccz:12} and between $-0.14$ and $-0.068~{\rm dex~kpc^{-1}}$ \citep{czc:11,krl:11,ksc:11,r:11,ccz:12}, respectively.

\begin{figure}
  \includegraphics[width=\hsize]{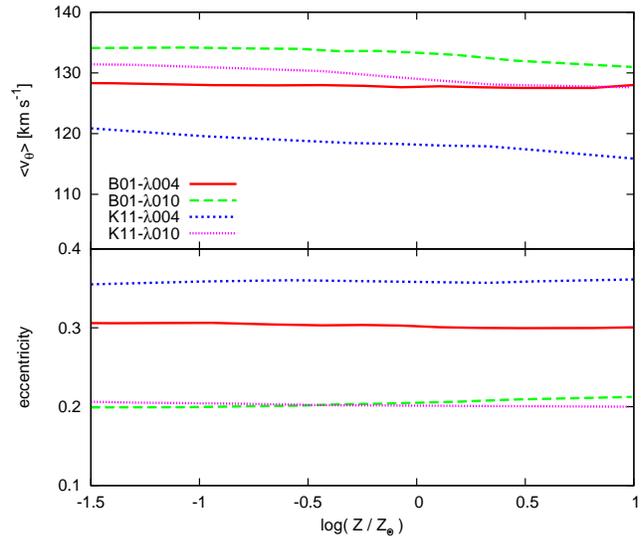}
  \caption{Orbital eccentricities and rotation velocities as functions of metallicities of stars in the region of $2<R/R_{\rm d}<3$ and $1~{\rm kpc}<|z|<4~{\rm kpc}$.}
  \label{ChemoDynamics}
\end{figure}
\citet{lba:11} and \citet{krl:11,kgw:13} have recently revealed that orbital eccentricity and rotation velocity seem to correlate with metallicity: $\epsilon$ decreases and $\overline{v_\theta}$ increases with metallicity for Galactic thick-disc stars. They found that these chemo-dynamical relations of the thick disc are distinct from those of the thin disc and can be used as proof of thick-disc formation scenarios. Fig. \ref{ChemoDynamics} shows these chemo-dynamical relations of thick-disc stars in our simulations. As clearly seen from the figure, neither $\epsilon$ nor $\overline{v_\theta}$ indicate correlation with metallicities. In connection to the above discussion, this result can be taken as a natural consequence of the weak gradients of metallicity and eccentricity shown in Fig. \ref{ChemProf} and the bottom panel of Fig. \ref{Ecce}. Accordingly, we can see that the clumpy disc scenario for thick-disc formation could not reproduce the correlations of $\epsilon$ and $v_\theta$ with metallicity observed in the Galactic thick disc since the violent mixing by massive clumps can erase the metallicity relations.

\subsection{Where do thick-disc stars form?}
\label{ClumpOriginMass}
It would be of great worth to discuss how and where the thick-disc stars formed in our simulations. In Fig. \ref{snap}, intense star formation can be seen in the clumps, and the inter-clump regions scarcely form stars. Notwithstanding, the galactic disc accounts for the greater fraction of the stellar mass in the final state. This implies that the giant clumps supply significant amount of stars to the disc. Giant clumps are totally or partly disrupted by tidal force while migrating toward the galactic centre. Besides, in merging with one another, the clumps scatter the inter-clump regions with stars. \citet{eem:09} have inferred that some fraction of the clump mass disperses into the inter-clump medium from their observations, and \citet[][hereafter BEE07]{bee:07} estimated that $\sim20$ per cent of disc mass is supplied from massive clumps in their simulations.

\begin{figure}
  \includegraphics[width=\hsize]{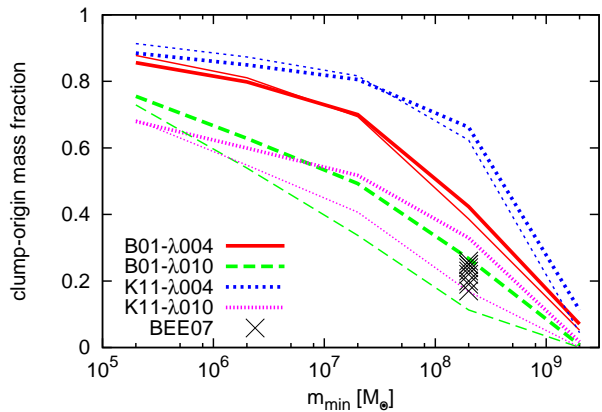}
  \caption{Mass fractions between stars released from clumps to the disc and the final stellar disc. The horizontal axis means minimum clump mass defined in the FOF method. The thick lines are for all disc stars, and the thin lines are for stars in the region of $2<R/R_{\rm d}<3$ and $1~{\rm kpc}<|z|<4~{\rm kpc}$ in the final snapshot. The $\times$-marks designate the results of BEE07; however, be aware of their different analysing technique (see the text).}
  \label{OriginMass}
\end{figure}

Taking advantage of the high mass-resolutions of our simulations, here we compute the mass fractions of stars released from clumps to discs. First, we identify stellar clumps in snapshots every $4~{\rm Myr}$ using the friends-of-friends (FOF) method, in which the linking lengths are set to $\left(m_{\rm s}/\rho_{\rm t}\right)^{1/3}$, where $m_{\rm s}$ is the mean mass of stellar particles and $\rho_{\rm t}$ is the stellar density at the radius where surface densities of a bulge and a disc become equal in the fittings of Fig. \ref{densities}. We found that the results do not depend very much on the linking length within a factor of approximately two. The minimum group (clump) mass $m_{\rm min}$ have to be given in the FOF algorithm, and we treat it as a parameter of the analyses. Here, we define FOF-grouped stars in the final snapshot to be bulge stars and the others to be disc stars. Next, we search for new stars born in the FOF clumps in each snapshot. If the new stars in the clumps are identified to be disc stars in the final snapshot, we consider these stars as the mass supply from the clumps to the disc.\footnote{Stellar mass loss due to SNe until the end of simulations is taken into account, i.e. the clump identification is performed in each snapshot, but the resultant mass fractions in Fig \ref{OriginMass} are computed with masses of stellar particles in the final snapshot.} Finally, we sum up the mass supply in all snapshots, estimate the mass ratio between the stars released from the clumps and the final disc at a given $m_{\rm min}$. The lowest-mass clump of $m_{\rm min}=2\times10^5{\rm M_\odot}$ consists of more than one hundred particles.

Fig. \ref{OriginMass} shows the results of the analysis as function of $m_{\rm min}$. We notify readers that this figure indicates mass-contribution cumulated from large to small clumps: steepness of the slopes at a certain $m_{\rm min}$ corresponds to contribution from the clumps having the mass. In the runs of B01-$\lambda$004 and K11-$\lambda$004, massive clumps of $\gsim10^8~{\rm M_\odot}$ largely contribute to disc formation, and the thick discs have almost the same composition as the entire discs. In B01-$\lambda$010 and K11-$\lambda$010, the contribution from smaller clumps is relatively larger than in the other runs, and all mass ranges contribute nearly equally, i.e. relatively straight profiles in the figure. We consider that this variance between the runs could be attributed to the difference of mass functions of clumps (see \S\ref{discuss_sim}). In any runs, it can be said that significant fractions of the thick-discs stars are born in the clumps, and disrupted clumps seem to be a major supplier of stars to galactic discs in our simulations. For comparison, we also show the results of simulations of BEE07 in the figure. Here we have to note that BEE07 have used different clump detection and analysis methods for the sake of fair comparison with observations: They deteriorated resolutions of their face-on density maps down to $200~{\rm pc}$ and searched for pixels with high contrast. BEE07 defined regions less extended than $3~{\rm kpc}$ and more massive than $2\times10^8{\rm M_\odot}$ as their clumps. Therefore, it should be kept in mind that fairness is not ensured in the comparison between our and BEE07 simulations in Fig. \ref{OriginMass}. Our results of B01-$\lambda$010 and K11-$\lambda$010 are nearly consistent with BEE07, but the runs of B01-$\lambda$004 and K11-$\lambda$004 indicate much larger mass contributions from massive clumps of $>2\times10^8{\rm M_\odot}$.

Fractional mass contribution from clumps of $10^7$--$10^9~{\rm M_\odot}$ thus seems to depend on the initial conditions of our simulations, therefore we could not assert that our simulations represent well the clumpy galaxies in the high-redshift Universe. In addition, feedback processes such as SNe and radiation pressure may be influential on the picture of disc formation and mass function of clumps \citep{g:12,pvt:13}. It is noteworthy, however, that the dynamical and chemical properties shown in the previous subsections were well consistent between the runs although Fig. \ref{OriginMass} showed the somewhat diverse results.

\section{Discussion}
\label{discuss}
\subsection{The validity and robustness of our simulations}
\label{discuss_sim}
First, we have to mention that numerical simulations for star-forming galaxies are still to be developed. Especially, implementation of feedback processes at the sub-grid level have long been debated \citep[e.g.][]{ssk:06,sbp:12,ds:12,akl:13}. Such feedback processes may be crucial for fate of giant clumps. \citet{n:96} has firstly pointed out the possibility that giant clumps may be transient objects if they are largely losing their mass by outflow. Recently, \citet{gnj:11} and \citet{nsg:12} observed that some clumps indeed seem to be subjected to such strong outflow. Afterwards, \citet{g:12} performed numerical simulations using a parametrised method for representing the superwinds and seemed to show that the clumps could not grow massive and would be disrupted in timescales shorter than their orbital periods. Some other observations also determined young ages of clumps in agreement with the ephemerality of clumps under the strong feedback \citep[e.g.][]{w:12,sss:12}. On the other hand, recent analytical studies and numerical simulations have taken the stellar feedback into account, and they seemed to show that giant clumps can accrete ambient gas to keep their masses even under the strong outflow (\citealt{kd:10,fkr:13,dk:13,bpr:13}; although see \citealt{hqm:12}). In this case, the clumps are expected to be long-lived and migrate toward the galactic centres. In addition, radial gradients of ages, specific star-formation rates and gas fractions of clumps were observed and could be taken as indicative of the longevity and the migration of clumps \citep{fsg:11,ggf:11,tkt:13II}; additionally, \citet[][]{mdc:13} confirmed the presence of these radial gradients in their simulations where their clumps are long-lived and form central bulges.

Because the stellar feedback processes have not been implemented in our simulation code yet, star formation rate would possibly be overestimated in the clumpy phase \citep[e.g.][]{sm:10,ds:12,akl:13}. The reader should be aware that our results may be sensitive to this issue. \citet{pvt:13} have performed $N$-body/SPH simulations using isolated disc models and shown how a mass function of clumps varies with strength of SN feedback between $0$ and $1.0\times10^{51}~{\rm erg}$ per SN. They found that a typical clump mass becomes more massive in a run with stronger SN, and the most frequent clump mass in the run with the strongest SN is $\sim0.5~{\rm dex}$ larger than in the case excluding the feedback. Thus, less massive clumps are disrupted due to the feedback, whereas massive ones can survive if they grow rapidly. \citet{bpr:13} also performed simulations with isolated galaxy models in which they took into account various feedback processes such as HII region photo-ionization, radiation pressure from young stars and non-thermal feedback from SNe. They demonstrated that clumps more massive than $\simeq10^8~{\rm M_\odot}$ can survive and keep their masses nearly constant even under the feedback stronger than in our study and \citet{pvt:13}; moreover, this result was confirmed in cosmological simulations (N. Mandelker et al. in prep.) In our study, Fig. \ref{OriginMass} indicates that more than 50 per cent of the thick-disc stars form in clumps more massive than $10^8~{\rm M_\odot}$ in the runs of B01-$\lambda$004 and K11-$\lambda$004 although the fractions are $\simeq20$ per cent in B01-$\lambda$010 and K11-$\lambda$010. Therefore, it could be expected that the thick-disc properties in B01-$\lambda$004 and K11-$\lambda$004 would be robust with respect to the feedback efficiency in simulations. On the other hand, the thick discs in B01-$\lambda$010 and K11-$\lambda$010 may not be formed from dissolved clumps if the other feedback processes are implemented. Moreover, bulge masses could also be affected by the feedback. If the feedback dissolves the majority of clumps, the bulges could not grow massive. In this sense, contribution by the bulges to potentials of entire galaxies may depend on the strength of the feedback, thereby the feedback efficiency may be influential on dynamical properties of disc stars via the bulge potentials.

Significance of the metallicity gradients would be affected by the strength of feedback. Cosmological simulations of \citet{gpb:13} have demonstrated that strong stellar feedback can wash out metallicity gradients in their discs although the gradients can be seen in the case of weak feedback. Our simulations would correspond to the case of weak feedback, therefore we could expect that the weak metallicity relations in this study would be due to the violent perturbation by the massive clumps. However, metallicity of the disc stars would become lower than in our simulations if the other feedback processes are implemented since star formation would be considerably suppressed and metal-enriched gas in clumps would be discharged by the feedback.

Our simulations are, in addition, not cosmological; our models are lacking continuous gas accretion which is expected to feed thin discs. The thin-disc formation could shorten the scale heights of the thick discs and might influence the velocity dispersions. In the cosmological context, smooth and cold gas streams are thought to penetrate a halo and keep a galaxy gas-rich until the streams are turned off in the high-redshift Universe \citep[e.g.][]{db:06,dbe:09}. As noted in \S\ref{RadialProfiles}, gas fraction in the clumps may affect some properties of the clump-origin bulges such as density profile and mass. In isolated models like our simulations, properties of giant clumps would be more or less controlled by the initial settings. For example, spin parameter is responsible for mass function of clumps: Small and large spin parameters of the initial conditions lead to high and low surface gas density of the discs. The wavelength at which dynamical instability first appears becomes longer in a disc with a higher surface density, which allows to form more massive clumps preferentially \citep{t:64,bt:08}. This inference can be seen in Fig.\ref{OriginMass}: In B01-$\lambda$010 and K11-$\lambda$010, more clumps of $\gsim10^8{\rm M_\odot}$ formed than in the runs of B01-$\lambda$004 and K11-$\lambda$004. As we showed in \S\ref{Results}, however, the kinematic and chemical properties of the thick discs in our simulations are qualitatively consistent between the runs and do not seem to depend very much on the initial conditions.

\subsection{The thick disc formation in clumpy galaxies}
\label{discuss_thick}
Here we inspect how realistic the scenario of thick-disc formation in clumpy galaxies is. As \citet{bem:09} showed, the thick-disc formation in clumpy galaxies has the advantages that constant thickness and old age of thick discs can be well reproduced in agreement with observations. These were confirmed in our simulations. In the scenarios of thin disc heating and radial migration, thick discs seem to form with scale heights increasing with galactocentric radius, i.e. flaring thick discs \citep{kbz:08,mfq:12}, although radial migration seems able to explain the relation between $\overline{v_\theta}$ and metallicity \citep{cls:12}. In our simulations, although the scale radii of the discs were contingent to the initial rotation of gas, the scale heights of $h_{\rm T}\simeq1~{\rm kpc}$ seem to be consistent with observed thick discs and almost independent of the initial conditions. Significant asymmetric drift, $\mathrm{d}\overline{v_\theta}/\mathrm{d}z$, was also confirmed in the runs of B01-$\lambda$004 and K11-$\lambda$004. Radial profiles of velocity dispersions except $\sigma_R$ were also nearly exponential. The pioneering work of \citet{lf:89} has observed exponential decrease of $\sigma_R$ and $\sigma_\theta$ for old disc stars with scale lengths similar to that of surface density, whereas \citet{dgk:11} observed the only significant radial gradient of $\sigma_z$ in the MW thick-disc stars. Moreover, we found that vertical profiles of velocity dispersions are almost isothermal in our simulations. In observations of the MW thick disc, \citet{mcm:12} have observed velocity dispersion profiles weakly increasing with height above the plane \citep[although see][]{sanders:12}; however, some observations indicated steep vertical gradients of $\sigma_R$ \citep{dgk:11} and $\sigma_z$ \citep{krl:11}. Distributions of orbital eccentricities in the entire thick discs in our simulations showed excellent agreement with observations of the MW thick disc. However, vertical dependence of the eccentricities cannot be reproduced. Thus, we can see that the comparisons of dynamical properties between our simulations and the observations do not indicate evident discrepancies, except the vertical variation of orbital eccentricities observed in the MW \citep{dkr:10}.

We found that metallicity gradients are quite weak in our simulations. In observations of the MW, thick-disc stars do not seem to have steep metallicity gradients in radial or vertical directions \citep[e.g.][]{abw:06,bao:11,r:11,krl:11,ccz:12}. However, our simulations did not show metallicity relations with orbital eccentricities and rotation velocities, which have recently been observed in the MW thick disc \citep{lba:11,krl:11,kgw:13}. We consider that this lack of chemo-dynamical relations is the most striking discrepancy between our simulations and the observations. We infer that disc stars are stirred by violent perturbation of giant clumps and such chemo-dynamical relations would be washed out in the clumpy phase. In the observations of \citet{lba:11}, their samples of thin- and thick-disc stars showed distinct chemo-dynamical relations, this is thought to reflect distinct formation processes of these discs. Thus, the chemical relations are expected to be susceptible to formation scenarios of discs, and the inconsistency between our simulations and the observations of the MW implies that the thick-disc formation scenario in clumpy galaxies may not be suitable for the MW. 

\citet{tkt:13II} have observed that 41 per cent of H$\alpha$ emitters of stellar masses from $10^9$ to $10^{11.5}~{\rm M_\odot}$ have clumpy morphologies at redshifts of 2.2 and 2.5, and the fraction of clumpy galaxies becomes the largest at the mass of $10^{10.5}~{\rm M_\odot}$ \citep[see also][]{mkt:14}. However, \citet{dln:13} observationally argued that MW progenitors would have stellar masses of $10^{9.5}$--$10^{10}~{\rm M_\odot}$ in this redshift range (see their fig. 1). From these observations, typical clumpy galaxies may be too massive to evolve into the MW-size galaxies. If it is the case, the thick-disc formation in a clumpy disc would not be suitable for the MW and smaller galaxies. 

It should be noted, however, that thick discs in external galaxies show a wide variety of kinematic properties, which is discussed to correlate with Hubble types and/or galaxy masses \citep[e.g.][]{yd:06,yd:08,cek:11}. Presumably, this may mean that the formation process of thick discs is not a single scenario and the diverse kinematics of the thick discs could be the consequence of the variety of formation history. Observational studies of chemo-dynamical properties of thick-disc stars in external galaxies are, unfortunately, far from sufficient. First, orbital eccentricity of each thick-disc star is not observable in these galaxies since three-dimensional position and velocity of a star are not accessible even if we could resolve its stellar image. However, it may be possible to observe the relation between metallicity and rotation velocity. \citet{yd:08} have observed some edge-on galaxies and obtained rotation curves at different height above disc planes. They found that relatively massive galaxies in their sample do not indicate significant vertical gradient of rotation velocities. On the other hand, their less massive galaxies show a wide range of behaviour such as counter-rotation between thin and thick discs, but they tend to indicate asymmetric drift more clearly than the massive galaxies. If altitudinal dependence of metallicity is observed in these galaxies, one can discuss whether they have a chemo-dynamical relation like the MW thick disc and whether it is a universal relation or not.

The authenticity of the clear distinction between the thin and thick discs is still under discussion, and the Galactic thick disc may be a superposition of relatively old components which are continuously connected to the thin disc \citep[e.g.][]{dt:93,bab:12I,brh:12III,bab:12II,sbr:13}. \citet{bab:12II} discussed that each component consisting of the Galactic disc is a simple structure with a single exponential density profile and isothermal kinematics. Recent observations by \citet{hdl:13}, however, indicated that the Galactic thick- and thin-disc stars have obviously distinguishable age-$\afe$ and -$\feh$ relations although the durations of their star formation somewhat overlap. Their result would prefer distinct formation processes between the thin and thick discs. They discussed the formation epoch of the thick disc to be the first $4$--$5~{\rm Gyr}$ of the Galaxy formation with starburst; this period corresponds to redshift of $z=2$--$3$. Thus, further investigation is needed to obtain more detailed and accurate information for the Galactic thick disc. In addition, observations for external galaxies are also essential to compare with the MW and inspect the universality and individuality of thick discs.

\section{Conclusions}
\label{conc}
In this study, we found that properties of thick discs are almost consistent among our runs with different initial conditions. Our simulations indicated that kinematic properties such as density and velocity profiles and distributions of orbital eccentricities are in acceptable agreement with Galactic and external galactic observations. However, altitudinal dependence of the eccentricities and metallicity relations with rotation velocities and eccentricities are not confirmed in our simulations, which are discrepant from recent observations of the Galactic thick disc. Therefore, we infer that the thick-disc formation scenario by clumpy discs would not be suitable at least for the MW.

\section*{Acknowledgments}

We would like to thank the referee for a careful reading of the manuscript and useful comments. We are grateful to Daisuke Kawata, Junichi Baba, Kohei Hattori, Shunsuke Hozumi, Ken-ichi Tadaki, Daniel Ceverino and Avishai Dekel for their helpful discussion. The numerical simulations reported here were carried out on Cray XT4 kindly made available by CfCA (Center for Computational Astrophysics) at the National Astronomical Observatory of Japan. SI acknowledges support from the Lady Davis Fellowship.

%\bibliography{references}

\appendix

\section[]{Robustness of the eccentricity determination}
\label{App}
Here we discuss how robust the eccentricity distributions presented in Fig. \ref{Ecce} are with respect to the modeling of galactic discs. The computation of stellar eccentricities have to model the mass distribution of the galaxies and might be significantly affected by our choice of the models. Especially, the disc model --- which is the only non-spherical component in the galaxy model --- may have a large impact on the eccentricity determinations.

In \S\ref{Eccentricity_disp}, our eccentricity determinations assumed the Miyamoto-Nagai discs \citep{my:75} for the sake of consistency with the analyses of \citet{lba:11}, and the structural parameters were chosen to fit with the circular velocity curves of the simulated galaxies. As shown in \S\ref{VerticalProfiles}, however, the density profiles of the stellar discs are fitted with the double-exponential models of equation (\ref{doubleExp}). In this appendix, we perform the same analyses for computing orbital eccentricities using the exponential models instead of the Miyamoto-Nagai discs. The radial and vertical forces, $K_R$ and $K_z$, caused by the exponential disc are given as

\begin{eqnarray}
  K_{R,i}(R,z) = -4\pi G\alpha_i\rho_{0,i}\int_0^\infty k\mathrm{d}kJ_1(kR)(\alpha_i^2+k^2)^{-\frac{3}{2}}&\nonumber\\
  \times\frac{\beta_ie^{-k|z|}-ke^{-\beta_i|z|}}{\beta_i^2-k^2}&
  \label{KR}
\end{eqnarray}
and
\begin{eqnarray}
  K_{z,i}(R,z) = -4\pi G\alpha_i\rho_{0,i}\int_0^\infty \mathrm{d}kJ_0(kR)(\alpha_i^2+k^2)^{-\frac{3}{2}}&\nonumber\\
  \times k\beta_i{\rm sign}(z)\frac{\beta_ie^{-k|z|}-ke^{-\beta_i|z|}}{\beta_i^2-k^2},&
  \label{Kz}
\end{eqnarray}
where $J_{0,1}$ are Bessel functions of the first kind, $\alpha_i\equiv R_i^{-1}$ and $\beta_i\equiv h_i^{-1}$ \citep{kg:89I}. The density models for the bulges and halos are the same as in \S\ref{DensityProfiles}. Although density distribution of gas seems different from the stellar discs in our simulations, we assume that the gas follows the thin-disc model.

\begin{figure}
  \includegraphics[width=\hsize]{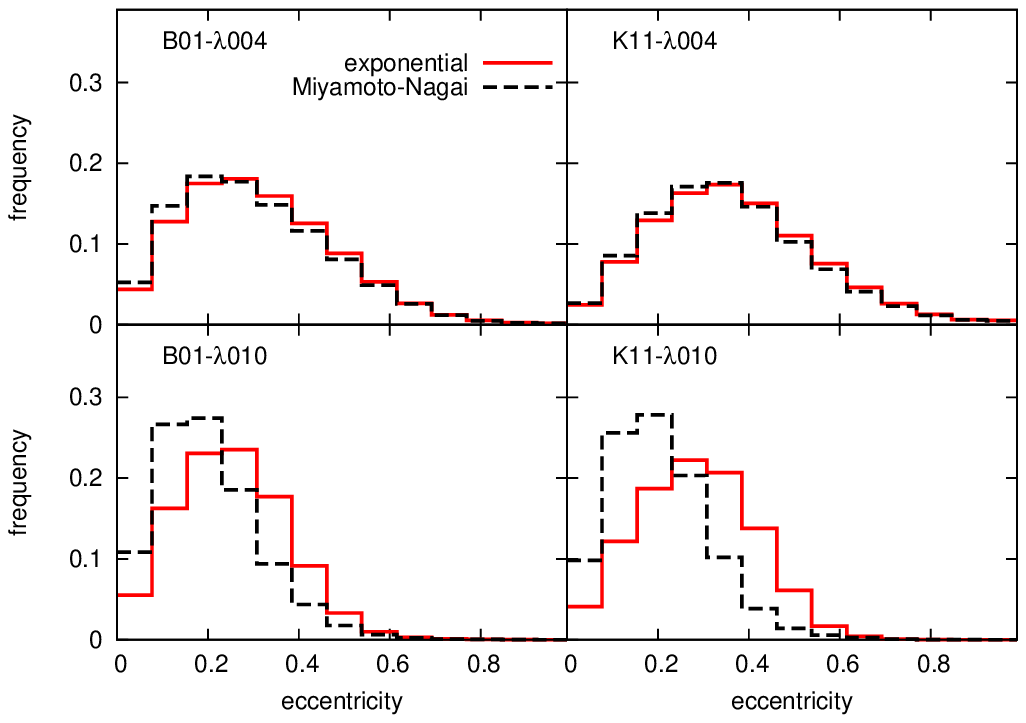}
  \includegraphics[width=\hsize]{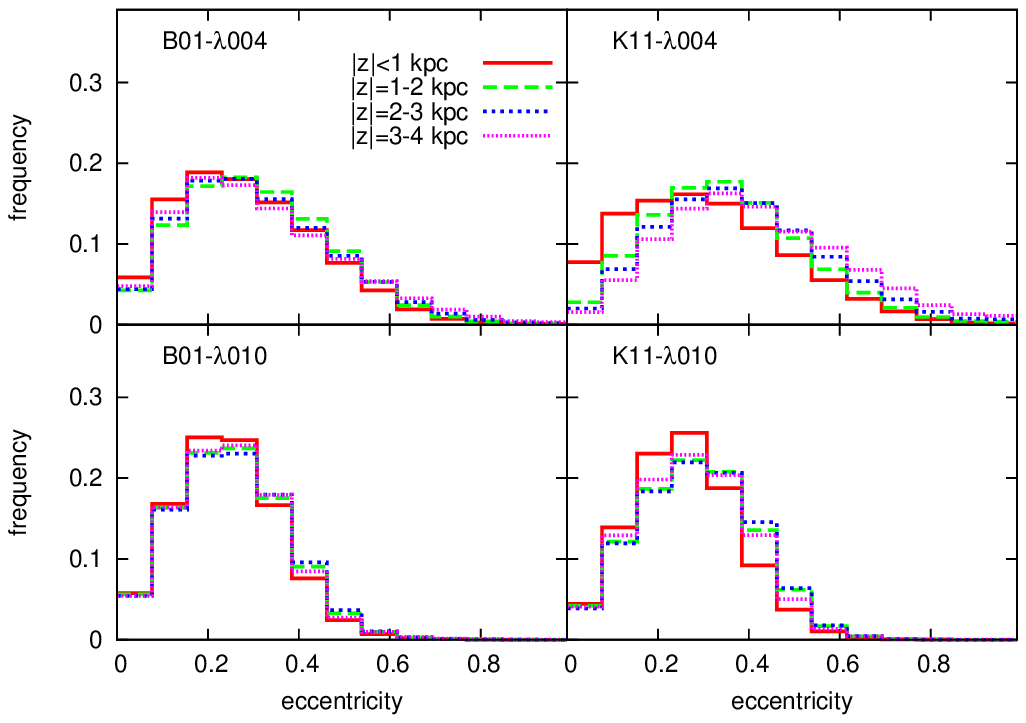}
  \caption{The same as Fig. \ref{Ecce} but using the double-exponential models for the thick and thin discs. In the top panel, the black dashed histogram indicates the result of Fig. \ref{Ecce} where the Miyamoto--Nagai disc model is used.}
  \label{Ecce_exp}
\end{figure}
Fig. \ref{Ecce_exp} shows the eccentricity distribution computed using the exponential disc models. In the top panel, we compare the eccentricities with the results in Fig. \ref{Ecce}. The eccentricity distributions are almost the same in the runs of B01-$\lambda$004 and K11-$\lambda$004. Although the distributions become to have a little higher averages and large dispersions in B01-$\lambda$010 and K11-$\lambda$010, our results do not change qualitatively. The bottom panel shows dependence of the distributions on distance from the planes. Systematic difference from the case using the Miyamoto--Nagai discs cannot be seen, and the distributions are still independent from the distance from the galactic planes. Hence, we can see that the eccentricity determinations in this study do not depend on the details of the galaxy models.

\end{document}